\documentclass[a4paper,12pt]{article}
\usepackage{epsfig}
\usepackage[body={156mm,257mm}]{geometry}

\begin{document}
%\textheight 25.7 cm
%\textwidth 15.6 cm
% Use the option doublespacing or reviewcopy to obtain double line spacing
% \documentclass[doublespacing]{elsart}

% if you use PostScript figures in your article
% use the graphics package for simple commands
% \usepackage{graphics}
% or use the graphicx package for more complicated commands
% \usepackage{graphicx}
% or use the epsfig package if you prefer to use the old commands
%\usepackage{epsfig}

% The amssymb package provides various useful mathematical symbols
%\usepackage{amssymb}

%\begin{frontmatter}



\centerline{\bf \large Intermediate phase in molecular networks and solid 
electrolytes}
$$ $$
\centerline{M. Micoulaut}

\centerline{Laboratoire de Physique Th{\'e}orique des Liquides}
\centerline{Universite Pierre et Marie Curie, Boite 121,}
\centerline{4, place Jussieu 75252 Paris Cedex 05 France}

$$ $$
\centerline{\bf Abstract}
{\small There is growing evidence that electronic and molecular
        networks present some common universal properties, among which the
        existence of a self-organized intermediate phase. In glasses,
the latter is
        revealed by the reversibility window obtained from complex
        calorimetric measurements at the glass transition. Here
        we focus on amorphous networks and we show how this intermediate
        phase can be understood from a rigidity percolation analysis on
        size increasing clusters. This provides benchmarks and guidance
        for an electromechanical analogy with high temperature super 
conductors.}
%\begin{keyword}

%\PACS 
%\end{keyword}
%\end{frontmatter}

% main text
\section{Introduction}
In the recent years, more and more studies devoted to the
understanding of the metal-insulator transition (MIT) have stressed the
possible existence of a self-organized electronic
phase\cite{r1,r3} that may explain the superconducting state. 
In these studies, in addition to the prediction or the measurement of
a critical dopant
concentration $n_{c1}$, it is noteworthy to stress that although the
mobile charge carrier concentration should tend to zero at the MIT 
and therefore also the interaction between the carriers, the
temperature $T_c$ should tend to zero, in contrast with current
observation. Moreover, a second anomalous behavior is observed at a
concentration $n_{c2}$ from the superconductive phase to the
non-superconductive Fermi liquid. This second transition is much more
abrupt and seems first order in character\cite{Meissner}. 
Illustrative examples of
these behaviors can be found in LSCO
($La_{2-x}Sr_xCuO_4$) from measurements of the filling factors by
Meissner volume (i.e. measuring the fraction of the material that is
super conductive)
and the transition temperature $T_c(x)$ with respect to Sr
concentration\cite{LASCO}. The nature of this intermediate phase 
in the dopant
region $n_{c1}<n<n_{c2}$ therefore still needs to be
understood. However, beyond the details of
modelling\cite{stripe}, a general
consensus emerge, which puts forward the idea of the
formation of modulated spatial patterns of mesoscopic length scale as
being responsable for HTSC\cite{r3,stripe}.
\par
Much more simpler examples of self-organization and the existence of
an intermediate phase can be provided by network glasses\cite{network}. Here,
instead of following the concentration of mobile carriers, focus is
made on elastic deformations of an amorphous network. Thus, instead of MIT,
percolative stiffness transitions\cite{percol} are obtained upon changing the
number of possible deformations which in turn can be related to the
connectivity of the molecular network. In order to describe these
deformations, one simple and elegant idea
is to translate the covalent interatomic valence forces of the atoms
in bond-stretching $n_c^\alpha$ and bond-bending mechanical constraints
$n_c^\beta$, using (mean-field)
Maxwell constraint counting\cite{Phil79}. A very particular point is reached 
when the number of
mechanical constraints $n_c=n_c^\alpha+n_c^\beta$ per atom equals the degrees of freedom per
atom in 3D, i.e. $n_c=3$ (corresponding to optimal glass formation)
\cite{Phil81}. 
This result has been obtained independently
from rigidity percolation by Thorpe\cite{Thorpe83} on disordered
networks, showing
that the number of zero 
frequency solutions (floppy modes) $f$ of the dynamical matrix of the
network equals $f=3-n_c$. The enumeration of constraints has been 
performed on a network 
of N atoms composed of $n_r$ atoms that are r-fold coordinated yielding 
$n_c^\alpha=r/2$ bond-stretching constraints and $n_c^\beta=2r-3$ bond-bending constraints
for a 
r-fold coordinated atom. The vanishing of the number of floppy modes
is 
obtained when the mean coordination number of the network reaches the 
critical value of 2.4, identified with an elastic stiffness
transition\cite{He}. This point defines the transition between a floppy network
which can be deformed without any cost in energy and a stressed rigid
network which has more constraints than degrees of freedom. At the
mean coordination number of 2.4, the network is isostatic. This new
phase transition has been studied with
considerable success in chalcogenides both in experiments and
numerical simulations\cite{chalco}.
Applications of rigidity 
theory have also been reported in various fields such as granular matter,
biology 
and computational science\cite{Traverse}. However, in the more recent
years it has
been shown that there 
should be two transitions\cite{PRL97} instead of the previously
reported single transition 
suggesting that the mean-field constraint counting alone may be
insufficient to 
describe completely the network change. Results on 
Raman scattering and measurements of the kinetics of the glass
transition using modulated differential scanning calorimetry
(MDSC) 
have indeed provided evidence\cite{PRL97,Selva} for two vibrational 
thresholds thus defining a stress-free intermediate phase
(isostatically rigid) in 
between the floppy and the stressed rigid phases, for which
independent 
evidence is obtained from numerical simulations\cite{JNCS00}.   
\par
   In this paper, we will construct a systematic approach that enables
us to perform constraint counting on size increasing structures thus
permitting to take into account medium range order effects such as small
rings. In doing this, we show that these small rings mostly control
the values of the critical coordination numbers and the width of the 
intermediate phase. The construction will
also show how isostatic regions and self-organization influence the
absolute magnitude of the width. The simplest case that can be build
up corresponds to Group IV chalcogenide glasses of the form
$B_xA_{1-x}$ with coordination numbers $r_A=2$ and $r_B=4$ defining  the mean
coordination number $\bar r=2+2x$. Size-increasing cluster 
approximations (SICA)
have been used in order to generate medium range order (MRO) from sets
of clusters on which
we have realized constraint counting. The results show two
transitions. A first one at which the number of floppy modes vanishes
and a second one (a ``stress transition'') beyond which stress in the
structure can not be avoided anymore out of ring structures. In
between, this defines an
almost stress-free network structure for which the rate of isostatic
regions can be computed. We will extend these results on solid
electrolytes for which percolative conductivity should be detected.
\section{Size increasing cluster approximations and constraint
counting}
Size increasing cluster approximations (SICA) have been first
introduced to describe the medium and
intermediate 
range order in amorphous semi-conductors\cite{Dina} but their
usefulness has been
also stressed for the description of the formation of
quasi-crystals\cite{quasi}
and fullerenes\cite{C60}. This approach emphasizes the rapid convergence of
significant MRO structures to a limit value when the size of the
considered clusters is increasing. Ideally, the infinite size cluster
distribution would yield the exact statistics of MRO in the
structure. The construction of these clusters is realized in Canonical
Ensemble with particular energy levels corresponding to bond creation
between short range order molecules (basic units) which are used as
building blocks from step l=1 (corresponding here to the reported
mean-field approach\cite{Phil79}) to arbitrary l. This construction should be
realized at the
formation of the network, when T equals the fictive temperature $T_f$
\cite{Galeener}. Since we expect to relate the width of the
intermediate phase with the ring fraction, we will restrict our present
study to Group IV chalcogenides of the form $Si_xSe_{1-x}$. For the
latter, there is strong evidence that at the stoichiometric
concentration x=0.33 a substantial amount of edge-sharing $SiSe_{4/2}$
tetrahedra\cite{Micol95}  can be found.
Therefore, we select basic units such as the $A_2$ (i.e. $Se_2$) chain
fragment and the stoichiometric $BA_{4/2}$ molecule
(e.g. $SiSe_{4/2}$). These basic units have
respective probabilities $1-p$ and $p=2x/(1-x)$, $x$ being the
concentration of the Group IV atoms.
We associate the creation of a chain-like $A_2-A_2$ structure (see
Table I) with an energy state of $E_1$, isostatic $A_2-BA_2$ bondings
with an energy gain of $E_2$ and corner-sharing (CS) and edge-sharing
(ES) $BA_{4/2}$ tetrahedra or any ring structure respectively with
$E_3$ and $E_4$. 
The probabilities of the different clusters have statistical weights
$g(E_i)$ which can be regarded as the degeneracy of the corresponding
energy level and correspond to the number of equivalent ways a cluster
can be constructed. Examples of statistical weights for the step $l=2$
are shown in Table I and in Refs\cite{Dina,quasi}.
\par
Due to the initial choice of the basic units, the energy $E_2$ will
mostly determine the probability of isostatic clusters since this
quantity is involved in the probability of creating the isostatic
$BA_4$ cluster (a $A_2-BA_{4/2}$ bonding). As a consequence, if we
have $E_2\ll E_1,E_3,E_4$, the network will be mainly isostatic in the
range of interest. 
\par
The calculation of the number of bond-bending $n_c^\alpha$ and
bond-stretching mechanical constraints $n_c^\beta$
per atom is performed on each cluster by Maxwell counting and
redundant constraints in ring structures are removed following the
procedure described by Thorpe\cite{Thorpe83}. It is obvious from the
construction that all the cluster probabilites will depend only on two
parameters (i.e. the factors $e_1/e_2$ and $e_3/e_2$) and eventually
$e_4/e_2$ if one considers the possibility of edge-sharing (ES) tetrahedra
or rings. One of the two factors has to be composition dependent since
a conservation law for the concentration of $B$ atoms $x^{(l)}$ has to be
fullfilled at any step $l$ of the construction\cite{Bray}:
\begin{equation}
\label{1}
x^{(l)}=x
\end{equation}
This means that either the fictive temperature $T_f$ or the energies
$E_i$ depend on $x$\cite{Galeener} but here only the $e_i(x)$
dependence is relevant for our purpose. 
Of course, when the step $l$ will increase, the number of isomers will
also increase among those also the number of different types of rings. 
The construction has been realized up to the step $l=4$ which already
creates clusters of MRO size. For each step $l$, we have determined either
$e_1/e_2$ or $e_3/e_2$ solving equ. (\ref{1}) and computed the total number of
constraints $n_c^l$:
\begin{eqnarray}
\label{2}
n_c^l={\frac {\sum_{i=1}^{N_l}n_{c(i)}p_i}{\sum_{i=1}^{N_l}N_ip_i}}
\end{eqnarray}
where $n_{c(i)}$ and $N_i$ are respectively the number of
constraints and the number of atoms of the cluster of size $l$ with 
probability $p_i$.
Once the factors become composition dependent, it is possible to find
for which concentration $x$ (or which mean coordination number $\bar
r$) the system
reaches optimal glass formation where the number of floppy modes
$f=3-n_c^l$ vanishes.
\section{Results}
Various possibilities can be studied within this framework.
The simplest case which can be investigated at the very beginning is
the random bonding case which is obtained when the cluster
probabilities $p_i$ are only given by their statistical weights (in
other words, the factors $e_i$ are set to one).
A single solution is obtained for the glass optimum point defined by
$f=0$ at  all SICA steps, in the mean coordination number range 
[2.231,2.275], slightly lower than the usual mean-field value of 2.4.
Since there is only one solution, no intermediate  phase in the case of random
bonding is found.
\par
Self-organization can be obtained by starting from a floppy
cluster of size $l$ (e.g. a chain-like structure made of a majority of
A atoms), and allowing the agglomeration of a new
basic unit onto this cluster to generate the cluster of size $l+1$ 
only if the creation of a stressed rigid region can be avoided
on this new cluster. The latter occurs when  two $BA_{4/2}$ basic
units are joined together.
With this rather simple rule, upon increasing $\bar r$ one
accumulates isostatic rigid regions on the size increasing clusters
because $BA_{4/2}$ units are only accepted in 
$A_2-BA_{4/2}$ isostatic bondings with energy $E_2$. Alternatively, we
can start from a
stressed rigid cluster
which exist at higher mean coordination number ($\bar r\leq 2.67$) and
follow the same procedure but in opposite way, i.e. we allow only
bondings which lead to isostatic rigid regions, excluding
systematically the possibility of floppy $A_2-A_2$ bondings.
In the case of self-organized clusters, the simplest case to be
studied is the case of dendritic clusters, when ring creation is
avoided. For an infinite size $l$, this would permit to
recover the results from Bethe lattice solutions or  
Random Bond Models\cite{RBM} for which rings are also exluded in the
thermodynamic limit\cite{Bethe}. A single transition for even $l$
steps at exactly the
mean-field value $\bar r=2.4$ is obtained whereas for the step $l=3$, there is
a sharp intermediate phase defined by $f=0$ (still at $\bar r=2.4$) and the 
vanishing of floppy regions (i.e. $e_1/e_2$ is zero) at $\bar
r=2.382(6)$. The probability of floppy, isostatic rigid and stressed
rigid clusters as a function of the mean
coordination number has been computed and shows
that the network is entirely isostatic at the point where $f=0$
(solid line, fig. 1). 
\par
The intermediate phase shows up if a certain amount of medium range
order (MRO) is allowed. This is realized in the SICA construction by
setting the quantity $e_4/e_2$ non zero, i.e. edge-sharing (ES) 
tetrahedra $BA_{4/2}$
leading to four-membered rings $B_2A_4$ can now be created at the growing
cluster steps. Two transitions are now obtained for
every SICA step. The first one is still located at $\bar r_{c1}=2.4$ (see
fig. 1). There,
the number of floppy modes $f$ vanishes. The second one is located
at $\bar r_{c2}$. When starting from a floppy network close to $\bar r=2$ and 
requiring self-organization (in this case, allowing only floppy or
isostatic bondings), this point corresponds to the network composition
beyond which stressed 
rigid regions outside of ring structures, which are created by the
dendritic connection of at least
two $BA_{4/2}$ units can not be avoided anymore. 
We call this point the ``stress transition''. We show in fig. 1
the $l=2$ result where $f=0$ at $\bar r_{c1}=2.4$ for different 
fractions of ES tetrahedra, defining an
intermediate phase $\Delta \bar r$. It is noteworthy to stress that the
first transition at $\bar r_{c1}$ corresponding to Phillips' glass optimum 
does not depend on the ES fraction, as well as the fraction of
stressed rigid clusters
in the structure. To ensure continuous deformation of the network when
B atoms are added and keeping the sum of the probability of floppy, isostatic
rigid and stressed rigid clusters equal to one, the probability of isostatic 
rigid clusters connects the isostatic solid line at $\bar r_{c2}$. 
Stressed rigid rings first appear in the region $\bar r_{c1}<\bar r<\bar r_{c2}$
while chain-like stressed clusters (whose probability is proportional
to $e_3$) occur only beyond the stress
transition, when $e_3\neq0$. This means that within this approach, when 
$\bar r$ is increased, stressed
rigidity nucleates through the network starting from rings, as ES
tetrahedra. It appears from fig. 1 that
the width $\Delta\bar r=\bar r_{c2}-\bar r_{c1}$ of
the intermediate phase increases with the fraction of ES. We have
represented the width as a function of the overall MRO fraction at the
rigidity transition in fig. 2 
which shows that $\Delta \bar r$ is almost an increasing function of the ES
fraction as seen from the result at SICA step $l=4$. Here, there is only a
small difference between allowing only four-membered rings (ES) (lower
dotted line) or
rings of all sizes (upper dotted line) in the clusters. Concerning the
nature of the structure in the intermediate phase, one can see from
fig. 1 and the insert of fig. 2 that the probability of isostatic clusters is
maximum in the window $\Delta \bar r$, and almost equal to 1 for
the even SICA steps, providing evidence that the molecular structure
in the window is almost 
stress-free. Furthermore, the latter can be extremely broad
if a high amount of edge-sharing is allowed. For a fraction of ES $\eta$
is close to one, molecular stripes emerge in the structure which are 
isostatic and every addition of new B atoms lowers
the amount of remaining A-chains by converting them in ES $BA_{4/2}$ stripes.
These molecular stripes are stress-free only if their size is infinite
\cite{Thorpe83}, which is the case for $\eta\simeq 1$ in the thermodynamic limit.
\section{Application to chalcogenides} 
As mentioned above, chalcogenide glasses are the first systems which
have been carefully studied in
the context of self-organization. 
SICA provides therefore a benchmark to check the results obtained. 
To be specific, Raman scattering has been used \cite{Bool1,Bool2} as a
probe to detect elastic thresholds in binary $Si_xSe_{1-x}$ and $Ge_xSe_{1-x}$
or ternary $Ge_xAs_xSe_{1-2x}$ glasses\cite{Euro00}. Changes in the germanium or
silicon corner sharing mode chain frequencies have
been studied with mean coordination number $\bar r$ of the glass network. These
frequencies exhibit a change in slope at
the mean coordination number $\bar r_{c1}=2.4$ and a first order jump
at the second transition $\bar
r_{c2}$. In germanium systems, the second transition is located around
the mean coordination number of $2.52$ whereas $\bar r_{c2}=2.54$ in Si
based systems. For both systems, a power-law behavior in $\bar
r-\bar r_{c2}$ is detected for $\bar r>\bar r_{c2}$ and the
corresponding measured exponent is very close to the one obtained in
numerical simulations of stressed rigid networks\cite{Tersoff}.
Moereover, a clear correlation between these results and the vanishing of the
non-reversing heat flow $\Delta H_{nr}$ (the part of the heat flow
which is thermal history sensitive) in MDSC measurements has been
shown\cite{Bool1,Bool2}.
\par
The SICA approach shows that the width $\Delta\bar
r$ of the intermediate phase increases mostly with the fraction of ES
tetrahedra. We stress that the width should converge to a lower limit
value of $\Delta \bar r$ compared to the step $l=2$, therefore one can
observe the shift downwards when increasing $l$ from 2 to 4. This
limit value is in principle attained for $l\to \infty$, or at least for much
larger steps than $l=4$ \cite{Micoul}. For Si-Se, $\Delta\bar r=0.14$ is
somewhat larger than for Ge-Se ($\Delta\bar r=0.12$) consistently with the
fact that the number of
ES is higher in the former\cite{Bool2}. 
\section{Application to solid electrolytes}
One interesting field of application of cluster construction and
constraint counting algorithms is the area of fast ionic conductors
(FIC)\cite{FIC1,FIC2}, which  has recently gained 
attention because of
potential applications of these solid electrolytes in all solid state
electrochemical devices and/or miniaturized systems such as solid state
batteries. This has been made possible by substitution of
more polarizable 
sulfur atoms, replacing the oxygen in usual oxide glasses and has led
to a substantial increase of the conductivity\cite{condux1}, as high as
$10^{-3}~\Omega^{-1}.cm^{-1}$, three orders of magnitude higher than the
conductivity of analogous oxide glasses\cite{condux2}.
Surpisingly, the extension of constraint theory from network glasses
(as $Si_xSe_{1-x}$) to FIC has been only reported for a few oxide
glasses\cite{Science,SSC}.  As a consequence, percolative effects have
not been studied so
far with the network change in FIC, although it seems fundamental for 
the understanding of the mobile alkali cations' motion. Since the
conductivity of the semi-conductor is proportional to the
mobility $\mu$,which in turn is related to the deformation of the 
network\cite{Barrio}, and the free carrier concentration $n_L$ by $\sigma=\mu n_L e$,
one may expect that the mobility in a floppy FIC should be substantially
higher compared
to the cation mobility in a stressed rigid network. This
means that the conductivity $\sigma$ should display some particular behavior in the
stress-free intermediate phase and at the two transitions.
\par
One can extend the SICA approach to the present case by considering
the simplest binary conducting glass, which is of the
form $(1-x)SiX_2-xM_2X$ with X an atom of Group VI (X=O,S,Se) and M an
alkali cation (M=Li,Na,K,\ldots). The local structure can be obtained from
NMR measurements and is mainly made of so-called $Q^3$ and $Q^4$
units\cite{Qn}. The former corresponds to the usual silica tetrahedron made of
one silicon and four Group VI atoms at the corner [e.g. $SiSe_{4/2}$] 
while the latter has
one oxygen atom ionically bonded to the alkali cation
[e.g. $SiSe_{5/2}^\ominus Na^\oplus$] (fig. 3).  
\par
Then, the probabilities can be evaluated for different steps of
cluster sizes following the procedure described in the previous
sections. It appears that the creation of a $Q^4-Q^4$ connection leads
to a stressed rigid cluster if the number of constraints is
computed, while the $Q^4-Q^3$ and $Q^3-Q^3$ connections yield
respectively isostatically stressed and floppy clusters. The SICA
results show again that the intermediate phase shows up if a non-zero
fraction of small rings is allowed in the structure, as displayed in 
fig. 4. 
Here in contrast with chalcogenides, the glass optimum
correponding to the
vanishing of the number of floppy modes respresents the upper limit of
the intermediate phase, which is consistent with the fact that one
starts from an almost stressed rigid molecular network at low modifier
concentration. Increase of the alkali content leads to an increase of 
floppiness. In the oxide system $(1-x)SiO_2-xM_2O$, the width
$\Delta x$ should be very small or zero since the fraction of ES in the
oxide systems is almost zero\cite{Galeener}. Still, percolative
effects are expected
at the concentration $x=0.2$ corresponding to the transition from
rigid to floppy networks. In sulfur and selenide glasses such as 
$(1-x)SiS_2-xNa_2S$, the width should be much broader because of the
existence of a high amount of edge-sharing tetrahedra in the $SiS_2$
or $SiSe_2$ base networks\cite{ES1}. In the sulfur base glass, $^{29}Si$ NMR
have shown that the rate of ES should be about 0.5, slightly higher
than in the selenide analogous system\cite{ES2}. As a result, one
should observe
a window of about $\Delta x=0.09$. Unfortunately, results on these systems
are only available for an alkali concentration
$x>0.2$\cite{Pradel}. However, in the
different silica based glasses, a rigidity transition has been 
observed\cite{Vaills} at
the concentration $x=0.2$ which should provide guidance for
forthcoming studies in this area.
\par
Finally, temperature effects should
be observable close to this transition. Since the concentration of alkali free
carriers $n_L$ depends on the temperature (the higher the temperature, the
higher $n_L$), an increase of the temperature $T$ should lead to a
decrease of the number of network constraints, the fraction of intact
bond-stretching constraints $n_c^\alpha$ of the alkali atom being
proportional to $1-n_L$. Consequently, a shift of the mechnical
threshold ($f=0$) to the higher concentrations should result from an
increase of T.
\section{Summary}
In this article, we have shown how stress change in  molecular systems
can be understood from the combination of cluster construction and
constraint counting. We have found that there is a single transition
from floppy to rigid networks in a certain number of structural
possibilities. An intermediate phase intervenes when a fraction of
non-zero molecular isostatic stripes are allowed, which makes this
new phase globally stress-free. The similarity of this new phase with
the superconducting filamentary phase in HTSC\cite{r3} but also in
biological networks\cite{Traverse} is striking and should
be worked out in more depth in the future.
\par
{\bf Acknowledgements}
It is pleasure to acknowldege ongoing discussions with R. Kerner,
G.G. Naumis, M.F. Thorpe and Y. Vaills. LPTL is Unit{\'e} de Recherche
associ{\'e}e au CNRS n. 7600.
% \appendix

% \section{}
% \label{}

\begin{table}
\begin{center}
\begin{tabular}{|c|c|c|c|c|}\hline
symbol&mechanical nature& &probability&$n_c$\\ \hline
 & & & & \\
$A_4$&floppy&\epsfig{figure=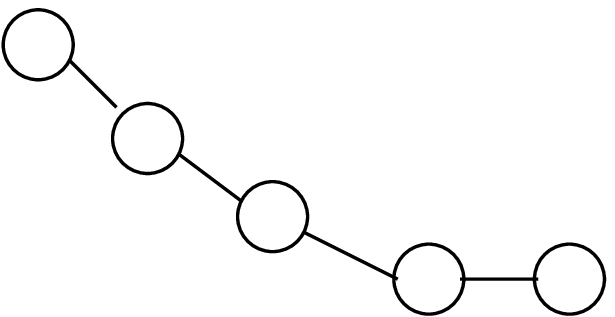,width=3cm}&$4(1-p)^2e_1$&2.0\\ \hline
 & & & & \\
$BA_4$&isostatic&\epsfig{figure=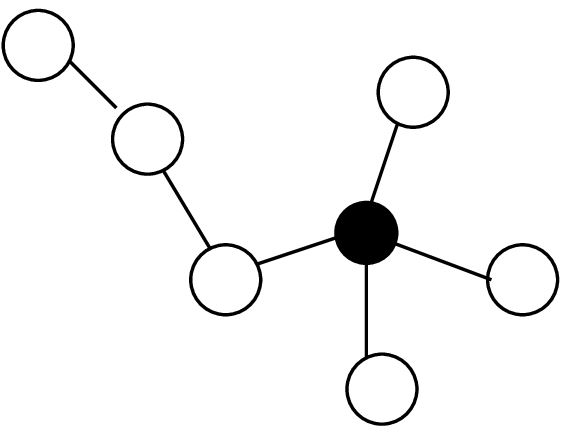,width=3cm}&$16p(1-p)e_2$&3.0\\
\hline
 & & & & \\
$B_2A_4$&stressed&\epsfig{figure=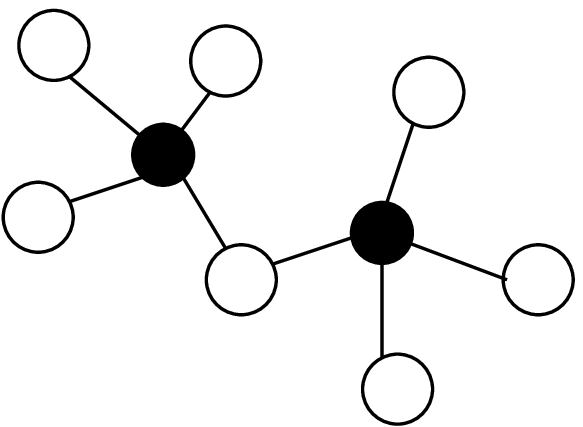,width=3cm}&$16p^2e_3$&3.67\\
\hline
 & & & & \\
$B_2A_4$&stressed&\epsfig{figure=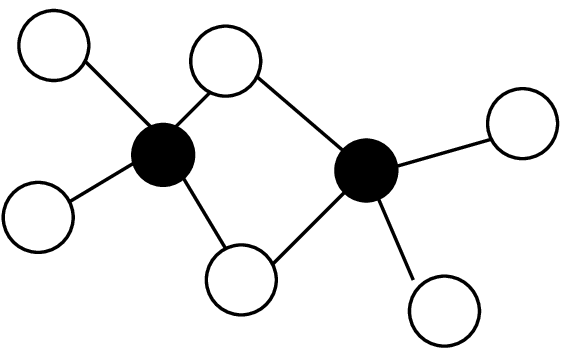,width=3cm}&$72p^2e_4$&3.33\\ \hline
\end{tabular}
\caption{The four clusters generated at step $l=2$ with their
unrenormalized probabilities and their number of mechanical
constraints. The factors $e_i$ are Boltzmann factors involving bond
energies and the fictive temperature, $e_i=exp[-E_i/T_f]$}
\end{center}
\end{table}
\newpage
\begin{figure}
\begin{center}
\epsfig{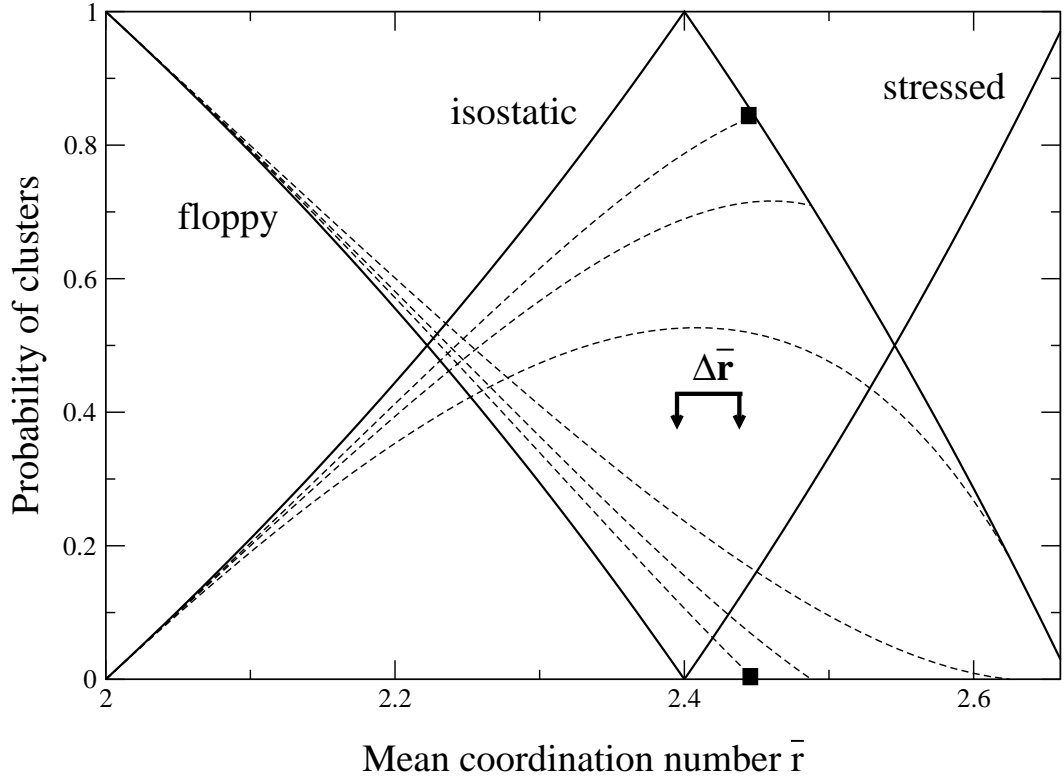}
\caption{\small Probability of finding floppy, isostatic rigid and stressed
rigid clusters as a function of mean coordination number $\bar r$ at SICA step
$l=2$ for different fractions of edge-sharing. The solid line
corresponds to the dendritic case where no rings are allowed
($e_4=0$). The broken lines correspond to situations whith a fraction
of ES at the stress transition of $\eta$=0.156, 0.290 and 0.818. The filled
squares indicate the point $\bar r_{c2}$ at which the stress
transition occurs in the case $\eta=0.156$.}
\end{center}
\end{figure}
\begin{figure}
\newpage
\begin{center}
\epsfig{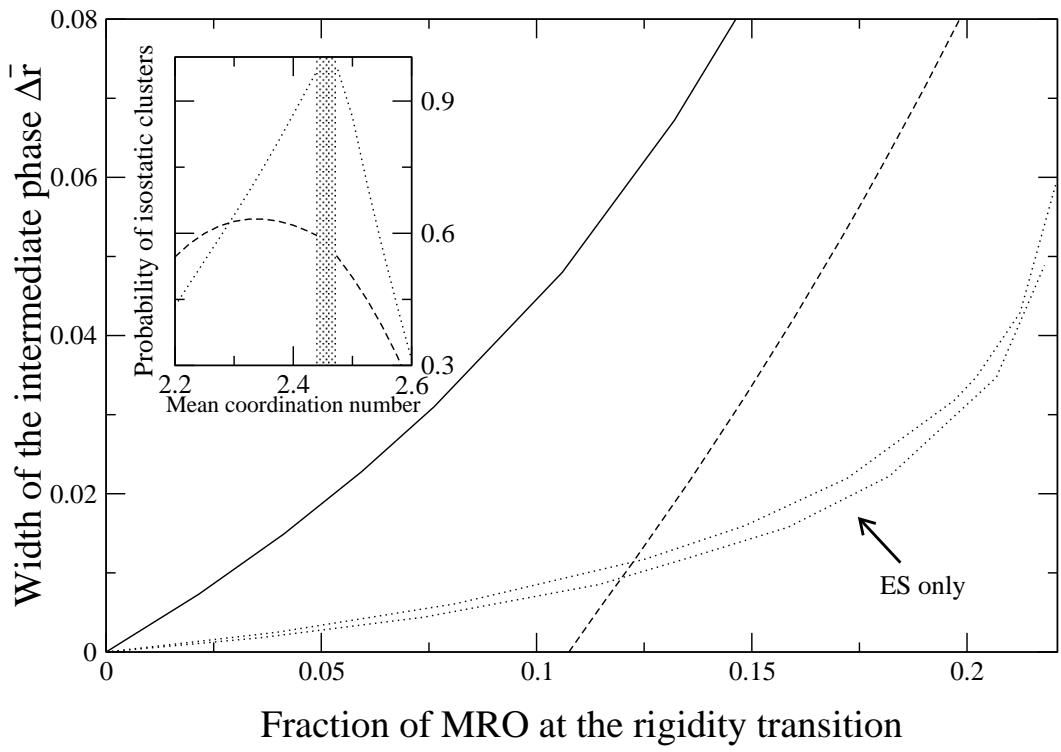}
\caption{\small Width of the intermediate phase $\Delta \bar r$ as a
function of the fraction of ring clusters at the rigidity transition 
for $l=2$ (solid line),
$l=3$ (dashed line) and $l=4$ (dotted lines). At step $l=2$, the
non-zero width comes only from the edge-sharing 
$BA_{4/2}$ tetrahedra. For $l>2$, different rings sizes
($4,6,8$) have been taken into account. The lower dotted line
corresponds to the intermediate phase at $l=4$ if only ES are allowed
as stressed rigid fragments. The upper dotted line is the same
quantity but allowing any ring structure. The insert shows the
probability of isostatic
clusters with mean coordination number $\bar r$ for $l=4$ (dotted
line) and $l=3$ (dashed line). The shaded region of $l=4$ represents the
intermediate phase.}
\label{f.3}
\end{center}
\end{figure}
\newpage
\begin{figure}
\begin{center}
\epsfig{figure=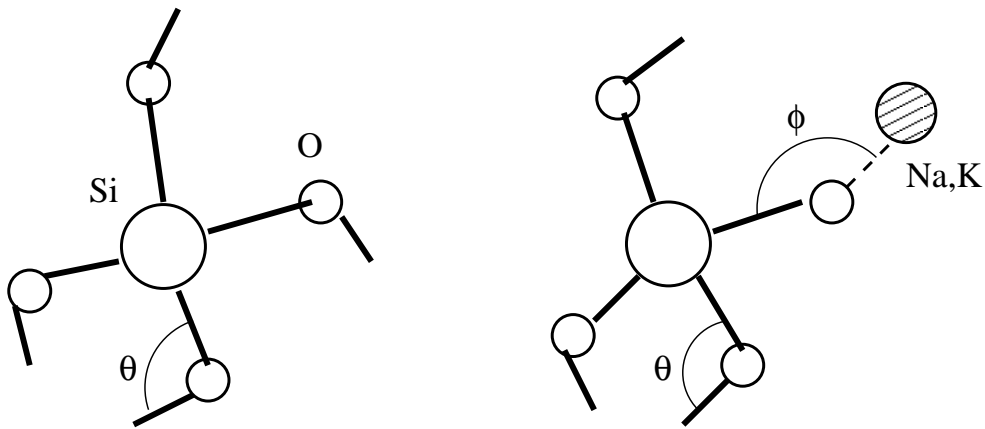,width=14cm}
\caption{\small The two local structures of the fast ionic conducting 
$(1-x)SiX_2-xM_2X$, ($X=O,S,Se; M=Li,Na,K$) glasses, denoted as $Q^4$
and $Q^3$ units, with respective probabilities $1-p$ and $p=2x/(1-x)$.}
\end{center}
\end{figure}
\newpage
\begin{figure}
\begin{center}
\epsfig{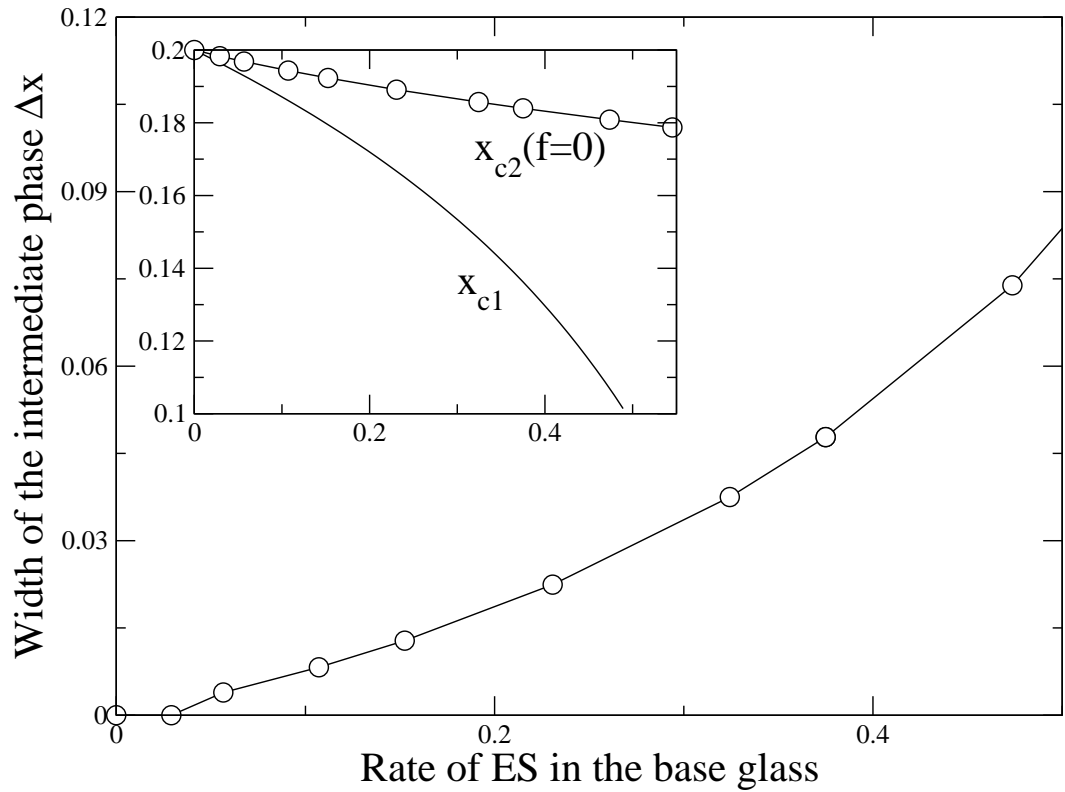}
\caption{\small Width of the intermediate phase $\Delta x$ as a function of the
rate of edge-sharing (ES) in the base glass. The insert shows the solutions
$x_{c1}$ and $x_{c2}$ obtained from SICA analyis, again as a function
of ES.}
\end{center}
\end{figure}

% notes:

% \bibitem{label2}
% If there is a note, it should come last:
% \bibitem{label3} \note
% \end{subbibitems}

\end{document}